\documentclass[preprint,floats,eqsecnum,aps,tightenlines,showpacs]{revtex4}
\usepackage{epsfig}
\usepackage{color}
\newcommand{\be}{\begin{equation}}
\newcommand{\ee}{\end{equation}}
\newcommand{\bea}{\begin{eqnarray}}
\newcommand{\eea}{\end{eqnarray}}
\newcommand{\nn}{\nonumber}

\newcommand{\ep}{i\epsilon}
\newcommand{\dq}{\frac{d^4q}{(2\pi)^4}}

\newcommand{\om}{\omega}
\begin{document}

\title{Solving the  Bethe-Salpeter equation for a pseudoscalar
meson in Minkowski space} 

\author{V.~\v{S}auli}
\affiliation{CFTP and Departamento de F\'{\i}sica,
Instituto Superior T\'ecnico, Av.\ Rovisco Pais, 1049-001 Lisbon,
Portugal, }
\affiliation{Department of Theoretical Physics,
Nuclear Physics Institute, \v{R}e\v{z} near Prague, CZ-25068,
Czech Republic}

\begin{abstract} 
A new method of solution of the Bethe-Salpeter equation for a pseudoscalar
quark-antiquark bound state is proposed. With the help of an integral
representation, the results are directly obtained in Minkowski space. 
Dressing of Green's functions is naturally taken into account, thus providing
the possible inclusion of a running coupling constant as well as quark
propagators. First numerical results are presented for a simplified ladder
approximation. 
\end{abstract}
\maketitle

\section{Introduction}

Among the various approaches used in meson physics, the formalism of
Bethe-Salpeter and Dyson-Schwinger equations (DSEs) plays a traditional and
indispensable role. The Bethe-Salpeter equation (BSE) provides a
field-theoretical starting point to describe hadrons as relativistic bound
states of quarks and/or antiquarks. For instance, the DSE and BSE framework has
been widely used in order to obtain nonperturbative information about the
spectra and decays of the whole lightest pseudoscalar nonet, with an 
emphasis on the QCD pseudo-Goldstone boson --- the pion \cite{PSEUDO}.  
Moreover, the formalism satisfactorily provides a window to the 'next-scale'
meson sector, too, including vector, scalar \cite{SCALARY} and excited mesons.
Finally, electromagnetic form factors of mesons have been calculated with this
approach for space-like momenta \cite{FORMFAKTORY}.
 
When dealing with bound states composed of light quarks, then it is unavoidable
to use the full covariant BSE framework. Nonperturbative knowledge of the
Green's function, which makes part of the BSE kernel, is  required. Very often,
the problem is solved in Euclidean space, where it is more tractable, as
there are no Green's function singularities there. The  physical amplitudes
can be then obtained by continuation to Minkowski space. Note that the
extraction of mass spectra is already a complicated task \cite{BHKRW2007}, not
to speak of an analytic continuation of Euclidean-space form factors.

When dealing with heavy quarkonia or mixed heavy mesons like $B_c$ 
(found at Fermilab by the CDF Collaboration \cite{BCmesons}),
 some simplifying  approximations are possible. Different
approaches have been developed to reduce the computational complexity of the
full four-dimensional (4D) BSE. The so-called instantaneous \cite{INSTA} and
quasi-potential approximations \cite{QUASI}  
can reduce the 4D BSE to a 3D equation in a Lorentz-covariant manner. In
practice, such 3D equations are much more tractable, since their resolution is
less involved, especially if one exploits the considerable freedom in
performing the 3D reduction. Also note that, contrary to the BSE in the ladder
approximation, these equations reduce to the Schr\"{o}dinger equation of
nonrelativistic Heavy-Meson Effective Theory and nonrelativistic QCD
\cite{HEAVY}. However, the interaction kernels of the reduced equations
often correspond to input based on economical phenomenological models, and the
connection to the underlying theory (QCD) is less clear (if not abandoned from
the onset).

In the present paper, we extend the method of solving the full 4D BSE,
originally developed for pure scalar theories \cite{NAKAN,KUSIWI,SAUADA2003}, 
to theories with nontrivial spin degrees of freedom. Under a certain
assumption on the functional form of Green's functions, we develop a method
of solving the BSE directly in Minkowski space, in its original manifestly
Lorentz-covariant 4D form. In order to make our paper as self-contained as
possible, we shall next supply some basic facts about the BSE approach to
relativistic mesonic bound states.

The crucial step to derive the homogeneous BSE for bound states is the
assumption that the bound state reflects itself in a pole of the four-point
Green's function for on-shell total momentum $P$, with $P^2=M_j^2$, viz.\
\be
G^{(4)}(p,p',P)=\sum_j\frac{-i}{(2\pi)^4}\frac{\psi_j(p,P_{os})
\bar{\psi_j}(p',P_{os})}{2E_{p_j}(P^0-E_{p_j}+\ep)}+\mbox{regular terms}\;, 
\ee 
where $E_{p_j}=\sqrt{\vec{p}\,{}^2+M_j^2}$ and $M_j$ is the (positive) mass of
the bound state characterized by the BS wave function $\psi_j$ carrying the
set of quantum numbers $j$. 
Then the BSE can be conventionally written in momentum space like
\bea
S_1^{-1}(p_+,P)\psi(p,P)S_2^{-1}(p_-,P)&&=-i\int\frac{d^4k}{(2\pi)^4}
V(p,k,P)\psi(p,P)\, ,
 \\
p_+&&=p+\alpha P \, ,
\nn \\
p_-&&=p-(1-\alpha)P \, ,
\nn
\eea
or, equivalently, in terms of BS vertex function $\Gamma$ as
\bea  \label{wakantanka}
\Gamma(p,P)&=&-i\int\frac{d^4k}{(2\pi)^4}V(p,k,P)S_1(k_+,P)
\Gamma(p,P)S_2(k_-,P) \, ,
\eea
where we suppress all Dirac, flavor and Lorentz indices, and $\alpha\in(0,1)$. 
The function $V $ represents  the two-body-irreducible interaction kernel, and
 $S_i$ ($i=1,2$) are the dressed propagators of the constituents. The free
propagators read
\be
S_i^0(p)=\frac{\not p+m_i}{p^2-m^2_i+\ep}.
\ee

Concerning solutions to the BSE (\ref{wakantanka}) for pseudoscalar mesons,
they have the generic form \cite{LEW}
\be \label{gen.form}
\Gamma(q,P)=\gamma_5[\Gamma_A+\Gamma_Bq.P\not\!q+\Gamma_C\not\!P+
\Gamma_D\not\!q\not\!P+  \Gamma_E\not\!P\not\!q] ,
\ee
where
the $\Gamma_i$, with $i=A,B,C,D,E$, are scalar functions of their arguments
$ P,q$. If the bound state has a well-defined charge parity, say 
${\cal{C}}=1$, then these functions are even in $q.P$, and furthermore
$\Gamma_D=-\Gamma_E$. 

As was already discussed in Ref.~\cite{MUNCZEK}, the dominant contribution to
the BSE vertex function for pseudoscalar mesons comes from the first term in
Eq.~(\ref{gen.form}). This is already true, at a 15\% accuracy level, for the
light pseudoscalars $\pi,K,\eta$, while in the case of ground-state heavy
pseudoscalars, like the $\eta_c$ and $\eta_b$, the contributions from the other
tensor components in Eq.~(\ref{gen.form}) are even more negligible.
Hence, at this stage of our Minkowski calculation, we also approximate our
solution by taking  $\Gamma=\gamma_5\Gamma_A$.

The interaction kernel is approximated by the dressed gluon propagator,
with the interaction gluon-quark-antiquark vertices taken in their bare forms.
Thus, we may write
\be \label{landau}
V(p,q,P)=g^2(\kappa) D_{\mu\nu}(p-q,\kappa)\gamma^{\nu}\otimes\gamma^{\mu} \, ,
\ee
where the full gluon propagator is renormalized at a scale $\kappa$. The
effective running strong coupling $\alpha_s$ is then related to $g$ through the
equations
\bea  \label{gluon}
g^2(\kappa)D_{\mu\nu}(l,\kappa)&&=
\alpha_s(l,\kappa)\frac{ P^T_{\mu\nu}(l)}{l^2+\ep}-\xi g^2(\kappa)
\frac{l_{\mu}l_{\nu}}{l^4+\ep}\, ,
\\
\alpha_s(q,\kappa)&&=\frac{g^2(\kappa)}{1-\Pi(q^2,\kappa)}\, ,
\nn \\
P^T_{\mu\nu}(l)&&=-g_{\mu\nu}+\frac{l_{\mu}l_{\nu}}{l^2}\, .
\nn 
\eea

From the class of $\xi$-linear covariant gauges, the Landau gauge $\xi=0$ will
be employed throughout the present paper. 

In the next section, we shall derive the solution for the dressed-ladder
approximation to the BSE, i.e., all propagators are considered dressed ones,
and no crossed diagrams are taken into account. The BSE for quark-antiquark
states has many times been treated in Euclidean space, even beyond the ladder
approximation. Most notably, the importance of dressing the proper vertices in
the light-quark sector was already stressed in Ref.~\cite{ACHJO}, so our
approximations are certainly expected to have a limited validity.
Going beyond the rainbow ($\gamma_{\mu}$) approximation is straightforward
but rather involved. (For comparison, see the Minkowski study of
Schwinger-Dyson equations published in Refs.~\cite{SAULIJHEP,SAULI2}),
the latter paper including the minimal-gauge covariant vertex instead of the
bare one). In the present paper, we prefer to describe the computational
method rather than carrying out a BSE study  with the most sophisticated 
kernel known in the literature.

The set-up of this paper is as follows. In Sec.~2 we describe the method of
solving the BSE. As a demonstration, numerical results are presented in 
Sec.~3. Conclusions are drawn in Sec.~4. The detailed derivations of the integral 
equation, that we actually solved numerically, are presented in the Appendices.

\section{Integral representation and solution of the BSE}

In this section we describe our method of solving the BSE in Minkowski space.
It basically assumes that the various Green's functions appearing in the
interaction kernel can be written as weighted integrals over the various
spectral functions (i.e., the real distribution) $\rho$.

More explicitly stated, the full quark and gluon propagators, the latter ones
in the Landau gauge, are assumed to satisfy the standard Lehmann
representation, which reads
\be \label{srforquark}
S(l)=\int_{0}^{\infty}d\om\frac{\rho_v(\omega)\not l+
\rho_s(\omega)}{l^2-\om+\ep}\,,
\ee
\be \label{srforgluon}
G_{\mu\nu}(l)=\int_{0}^{\infty}d\om\frac{\rho_g(\omega)}{l^2-\om+\ep}
P^T_{\mu\nu}(l) \, , 
\ee
where $\rho$ is a real istribution.
 Until now, with certain limitations, the
integral representations ~(\ref{srforquark}) and (\ref{srforgluon}) have been used for the nonperturbative evaluation
of Green's functions in various models \cite{SAFBP}. However, we should note
here that the true analytic structure of QCD Green's functions is not reliably
known (also see Refs.~\cite{ALKSME,FISHER1,SABIA}), which studies suggests the tructure given by ~(\ref{srforquark}) and (\ref{srforgluon}) is not sufficient if not excluded. In this case, the lehmann representation or perhaps the ussage of real $\rho$ in the integral representation ~(\ref{srforquark}) and (\ref{srforgluon}) can be regarded as an analyticized approximation of the true
quark propagator. The complexification of $rho$ within the complex integration path is one of the straightforward  and  questionable generalization \cite{ARRBRO}. The general question of the existence of Lehamnn represintation in QCD is beyond the scope of presented paper and we do not discussed the problem furthermore.

Furthermore, we generalize here the idea of the Perturbation Theory Integral
Representation (PTIR) \cite{NAKAN}, specifically for our case. The PTIR
represents a unique integral representation (IR) for an $n$-point Green's function
defined by an $n$-leg Feynman integral. 
The generalized PTIR formula for the $n$-point function in a theory involving
fields with arbitrary spin is exactly the same as in the original scalar theory
considered in Ref.~\cite{NAKAN}, but the spectral function now acquires a
nontrivial tensor structure. Let us denote such a generalized weight function by
$\rho(\alpha,x_i)$.  Then, it can be clearly decomposed into the sum 
\be
\rho(\alpha,x_i)_{\mbox{\scriptsize scalar theory}}\rightarrow \sum_j
\rho_j(\alpha,x_i){\cal{P}}_j  ,
\ee
where $\alpha,x_i$ represent the set of spectral variables, and $j$ runs over
all possible independent combinations of Lorentz tensors and Dirac matrices
$P_j$. The function $\rho_j(\alpha,x_i)$ just represents the PTIR weight
function of the $j$-th form factor (the scalar function by definition), since
it can obviously be written as a suitable scalar Feynman integral. Leaving
aside the question of (renormalization) scheme dependence, we refer the reader
to the textbook by Nakanishi \cite{NAKAN} for a detailed derivation of the
PTIR. The simplest examples of such "generalized" integral representations corresponds with Lehmann
representations for spin half ~(\ref{srforquark}) and spin one propagators  (\ref{srforgluon}). 
 
Let us now apply our idea to the pseudoscalar bound-state vertex function
keeping in mind that the singularity structure (given by the denominators)
of the r.h.s.\ of the  BSE is the same as in the scalar models studied in
Refs.~\cite{KUSIWI,SAUADA2003}, the appropriate IR for the pseudoscalar
bound- state vertex function $\Gamma_A(q,P)$ should read
\be  \label{repr}
\Gamma_A(q,P)=\int_{0}^{\infty} d\om \int_{-1}^{1}dz 
\frac{\rho_A^{[N]}(\om,z)}
{\left[F(\om,z;P,q)\right]^N}\, ,
\ee
where we have introduced a useful abbreviation for the denominator of the
IR~(\ref{repr}), viz.\
\be \label{efko}
F(\om,z;P,q)=\om-(q^2+q.Pz+P^2/4)-\ep \, ,
\ee
with $N$ a free integer parameter. 

Substituting the IRs~(\ref{repr}), (\ref{srforgluon}), (\ref{srforquark}) into
the r.h.s.\ of the BSE~(\ref{wakantanka}), one can analytically integrate over
the loop momenta. Assuming the uniqueness theorem \cite{NAKAN}, we should
arrive at the same IR~(\ref{repr}), because  of the r.h.s.\ of the
BSE~(\ref{wakantanka}). The derivation is given in Appendix A for the cases
$N=1,2$.

In other words, we have converted the momentum BSE (with a singular kernel)
into a homogeneous two-dimensional integral equation for the real weight
function $\rho_A^{[N]}(\om,z)$, i.e.,
\be \label{madrid}
\rho^{[N]}_A(\tilde{\om},\tilde{z})=
\int_{0}^{\infty} d\om \int_{-1}^{1}dz
V^{[N]}(\tilde{\om}, \tilde{z};\om,z)\rho^{[N]}_A(\om,z) ,
\ee
where the kernel $V^{[N]}(\tilde{\om}, \tilde{z};\om,z)$ is a regular
multivariable function.

The kernel $V^{[N]}$ also automatically supports the domain $\Omega$ where
the function $\rho^{[N]}_A(\om,z)$ is nontrivial. This domain is always 
smaller then the infinite strip $[0,\infty)\times[-1,1]$, as is explicitly
assumed by the boundaries of the integrals over $\om$ and $z$.
For instance, with the simplest kernel parametrized by a free gluon
propagator and constituent quarks of mass $m$, we get for the flavor-singlet
meson $\rho^{[N]}_A(\om,z)\neq 0$ only if $\om>m^2$.

In our approach, to solve the momentum BSE in Minkowski space is equivalent
to finding a real solution to the real integral equation~(\ref{madrid}). No
special choice of frame is required. If one needs the resulting vertex
function, can be obtained by numerical integration over $\rho_N$ in an
arbitrary reference frame.

%%%%%%%%%%%%%%%%%%%%%%%%%%%%%%%%%%%%%%%%%%%%%%%%%%%%%%%%%%%%%%%%%%%%

\section{ Numerical Results}

In this section we discuss the numerical solution of the BSE with various 
interaction kernels. For that purpose, we shall vary the coupling strength 
as well as the effective gluon mass $m_g$. We are mainly concerned with the
range of binding energies that coincide with those of heavy quarkonia, which
systems we shall study in future work. Moreover, we take a discrete set of
values for the mass $m_g$, such that it runs from zero to the value of the
constituent quark mass. These values are expected to be relevant for the case
of a true gluon propagator (when $m_g$ is replaced by the continuous spectral
variable $\om$ (\ref{srforgluon})). Thus, in each case, the corresponding
gluon density is  $\rho_g(c)=N_g\delta(c-m^2_g)$, which specifies the kernel
of the BSE to be (in the Landau gauge)
\be  \label{gluon2}
V(q-p)=g^2
\frac{-g_{\mu\nu}+\frac{(q-p)_{\mu}(q-p)_{\nu}}{(q-p)^2}}
{(q-p)^2-m_g^2+\ep}
\gamma^{\nu}\otimes\gamma^{\mu}
\ee
where the prefactor (including the trace of the color matrices) is simply
absorbed in the coupling constant. For our actual calculation, we use the bare
constituent propagator $S_i(p_i)$ with heavy quark mass $M\equiv m$ (see Appendix A
for this approximation).

Firstly, we follow the standard procedure: after fixing the bound-state mass
($\sqrt{P^2}$), we look for a solution by iterating the BSE for a spectral
function with fixed coupling constant $\alpha=g^2/(4\pi)$. 
Very similarly to the scalar case \cite{SAUADA2003}, the choice $N=2$ for the
power of $F$ in the IR of the bound-state vertex function is the preferred one.
This choice is a reasonable compromise between on the one hand limiting
numerical errors and on the other hand avoiding the computational obstacles
for high $N$. Here we note that using $N=1$ is rather unsatisfactory
(comparing with the massive Wick-Cutkosky model), since then we do not find any
stable solution for a wide class of input parameters $g$, $m_g$. In contrast,
using the value $N=2$ we obtain stable results for all possible interaction
kernels considered here. This includes the cases with vanishing $m_g$, which
means that the numerical problems originally present in the  scalar models
\cite{SAUADA2003} are fully overcome here. The details of our numerical
treatment are given in Appendix B. 

As is more usual in the nonrelativistic case, we fix the coupling constant
$\alpha=g^2/(4\pi)$ and then look for the bound-state mass spectrum. We find
the same results in either case, whether $P$ or $\alpha$ is fixed first, 
noting however that in the latter case the whole integration in the kernel
$K$ needs to be carried out in each iteration step, which makes the problem
more computer-time consuming.

\begin{figure}
\centerline{  \mbox{\psfig{figure=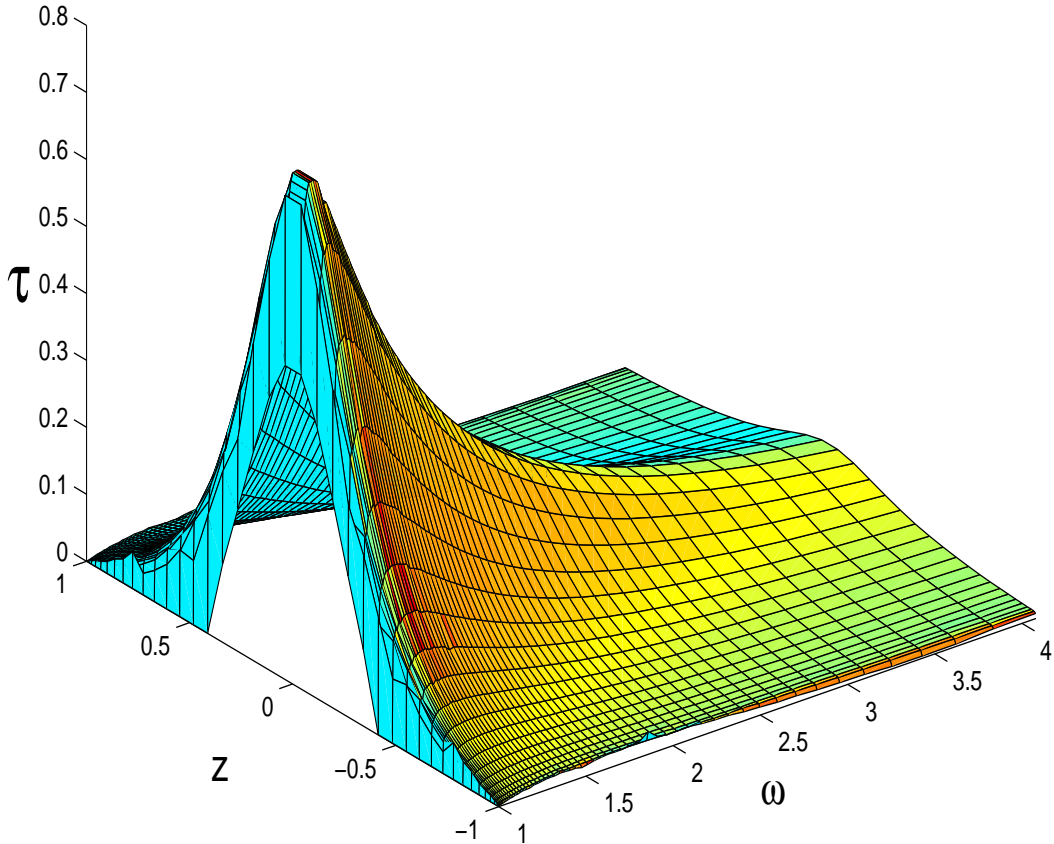,height=14.0truecm,
width=14.0truecm,angle=0}} }
\caption[99]{\label{figBSE} The rescaled weight function 
$\tau=\frac{\rho^{[2]}(\om,z)}{\om^2}$ for the following model parameters:
$\eta=0.95$, $m_g=0.001M$, $\alpha_s=0.666$; the small mass $m_g$ 
 approximates the one-gluon-exchange interaction kernel. }  
\end{figure}

The obtained solutions for varying $\alpha$ and mass $m_g$, with
a fixed fractional binding $\eta=\sqrt{P^2}/(2M)=0.95$, are given in Table~1.
If we fix the gluon mass at $m_g=0.5$ and vary the fractional binding $\eta$,
we obtain the spectrum of Table~2.

%Table 1
\begin{center}
\small{\begin{tabular}{|c|c|c|c|c|}
\hline \hline
$m_g/m_q $ &   $10^{-3}$  & 0.01  & 0.1 & 0.5   \\
\hline 
 $\alpha$ & 0.666 & 0.669  & 0.745 & 1.029  \\
\hline \hline
\end{tabular}}
\end{center}

\begin{center}
TABLE 1. Coupling constant $\alpha_s=g^2/(4\pi )$  for
several choices of $ m_g/M$,
with given binding fraction $\eta=\sqrt{P^2}/(2M)=0.95$.
 \end{center}

%Table 2
\begin{center}
\small{\begin{tabular}{|c|c|c|c|c|}
\hline \hline
$\eta: $ &0.8 &  0.9  & 0.95 & 0.99   \\
\hline 
 $\alpha$ &1.20 & 1.12 & 1.03  & 0.816  \\
\hline \hline
\end{tabular}}
\end{center}

\begin{center}
TABLE 2. Coupling $\alpha_s=g^2/(4\pi )$ as a 
function of binding fraction $\eta=\sqrt{P^2}/(2M)$, for  
exchanged massive gluon with $m_g=0.5M$. 
\end{center}

For illustration, the  weight function $\tilde{\rho}^{[2]}$  is
displayed in Fig.~\ref{figBSE}.

\section{ Summary and  Conclusions}
 
The main result of the present paper is the development of a technical
framework  to solve the bound-state BSE in Minkowski
space. In order to obtain the spectrum, no preferred reference frame is
needed, and the wave function can be obtained in an arbitrary frame
--- without numerical boosting --- by a simple integration of the weight
function.

The treatment is based on the usage of an IR for the
Green's functions of a given theory, including the bound-state vertices
themselves. The method has been explained and checked numerically on the
samples of pseudoscalar fermion-antifermion bound states. It was shown
that the momentum-space BSE can be converted into a real equation for a
real weight functions $\rho$, which is easily solved numerically.
The main motivation of the author was to develop a practical tool respecting 
selfconsistency of DSEs and BSEs. Generalizing this study to other mesons,
such as vectors and scalars, and considering more general flavor or isospin
structures, with the simultaneous improvement of the approximations
(correctly dressed gluon propagator, dressed vertices, etc.), will be an
essential step towards a fully Lorentz-covariant description of a plethora of
transitions and form factors in the time-like four-momentum region. 

\

\
 
{\Large{ \bf Acknowledgments}}

\

I would like to thank George Rupp for his careful reading of the manuscript. 
 
%%%%%%%%%%%%%%%%%%%%%%%%%%%%%%%%%%%%%%%%%%%%%%%%%%%%%%%%%%%%%%%%%%%%%%%%%%%%%%%%%%%%%%%%%%%%%%%%%%%%%%%%%%%%%%%%%%%%
\appendix
\section{Kernel Functions} 

With the Dirac indices explicitly written out, the BSE for quark-antiquark
bound states reads
\begin{equation} \label{BSE}
\Gamma(q,P)_{\om\rho}=i\int \dq
S(q+P/2)_{\beta\gamma}\Gamma(q,P)_{\gamma\gamma'}
S(q-P/2)_{\gamma'\beta'} V_{\om\beta\beta'\rho}(q,k;P) ,
\end{equation}
where the Lorentz indices of the vertex function have not been specified.

In our approximation, the IR for a pseudoscalar
bound-state vertex function is
\be  \label{reprnul}
\Gamma(q,P)^{\alpha\beta}=\gamma_5^{\alpha\beta}\int_{0}^{\infty} d\om
\int_{-1}^{1}dz \frac{\rho^{[N]}(\om,z)}
{\left[F(\om,z;P,q)\right]^N} ,
\ee
and the generalized kernel for the BSE in the ladder approximation has the form
\be \label{irforker}
V_{\alpha\beta\gamma\delta}=g^2\int_{0}^{\infty}d c
\frac{-g_{\mu\nu}+\frac{(q-p)_{\mu}(q-p)_{\nu}}{(q-p)^2}}
{(q-p)^2-c+\ep}\rho_g(c)
\gamma^{\nu}_{\alpha\beta}\gamma^{\mu}_{\gamma\delta},
\ee
where the indices $\alpha,\beta,\gamma,\delta$ and $\mu,\nu$ stand for
the appropriate Dirac and Lorentz structures, respectively.
Moreover, we use the IR for all functions entering the BSE, including
the vertex (\ref{reprnul}), the kernel (\ref{irforker}), and the propagators
$S(q_\pm)$ (\ref{srforquark}).

For the purpose of brevity, we shall use the following abbreviation for the
prefactor:  
\be
\int_{\cal{S}}\equiv
-3\int_{0}^{\infty} d\om \int_{-1}^{+1}dz \int_0^{\infty}dc
 \int_0^{\infty}da\int_0^{\infty} db  \rho^{[N]}(\om,z) g^2\rho_g(c) \, .
\ee
With this convention, the BSE can be written as

\bea\label{main}
\int d\tilde{\om} d\tilde{z} \frac{\rho^{N}(\tilde{\om},\tilde{z})}
{\left[F(\tilde{\om},\tilde{z};p,P)\right]^N}
=i \int_{\cal{S}} \int\dq 
\frac{\rho_v(a)\rho_v(b)(q^2- P^2/4)-\rho_s(a)\rho_s(b)}
{\left[F(\om,z;q,P)\right]^N D_1 D_2 D_3} \, ,
\eea
where the trace is taken over the Dirac indices (after multiplying by
$\gamma_5$), and where
\bea
D_1&=&(q+P/2)^2-a+\ep \, ,
\nn \\
D_2&=&(q-P/2)^2-b+\ep \, ,
\nn \\
D_3&=&(q-p)^2-c+\ep \, .
\eea
In the following, we shall transform the r.h.s.\ of the BSE~(\ref{main}) into 
its IR, i.e., the l.h.s.\ of Eq.~(\ref{main}).

As a first step, we use the algebraic identities
\be
\frac{q^2}{F(\om,z;q,P)}=
\frac{\om-q.Pz-P^2/4}{F(\om,z;q,P)}-1 \, ,
\nn
\ee
\be
\frac{q.P}{D_1D_2}=\frac{1}{2}\left(\frac{1}{D_1}-\frac{1}{D_2}
+\frac{b-a}{D_1D_2}\right) \, ,
\ee
which gives us for the r.h.s.\ of Eq.~(\ref{main})
\bea \label{prava2}
&&i \int_{\cal{S}} \int\dq
\left\{\frac{-\frac{z}{2}\rho_v(a)\rho_v(b)}{\left[F(\om,z;q,P)\right]^N D_3}
\left(\frac{1}{D_1}-\frac{1}{D_2}\right)\right.
\nn \\
&+&\frac{(\om-P^2/2+\frac{a-b}{2}z)\rho_v(a)\rho_v(b)-\rho_s(a)\rho_s(b)}
{\left[F(\om,z;q,P)\right]^N D_1 D_2 D_3}
-\frac{\rho_v(a)\rho_v(b)}{\left[F(\om,z;q,P)\right]^{N-1}D_1 D_2 D_3} .
\eea
Furthermore, we employ Feynman parametrization, starting with the first term
in expression~(\ref{prava2}), which yields
%page5pi
\bea \label{match}
&&\frac{1}{\left[F(\om,z;q,P)\right]^N }
\left(\frac{1}{D_1}-\frac{1}{D_2}\right)=
(-)^N\int_0^1dx\frac{\Gamma(N+1)x^{N-1}}
{\Gamma(N)}
\nn \\
&&\left\{\left[q^2+q.P(zx-(1-x))+P^2/4-\om x-a(1-x)\right]^{-N-1}\right.
\nn \\
&&-\left.\left[q^2+q.P(zx+(1-x))+P^2/4-\om x-b(1-x)\right]^{-N-1}\right\}\, .
\eea 
Substituting Eq.~(\ref{match}) back into expression~(\ref{prava2}),
using the Feynman variable $y$ so as to match the scalar propagator $D_3$,
and integrating over the four-momentum $q$, we get for the first line in 
expression~(\ref{prava2}) 
\bea   \label{racek}
&&(-)^{N-1}\int_{\cal{S}}\frac{\rho_v(a)\rho_v(b)}{2(4\pi)^2}
\int_0^1 dx\int_0^1dy y^Nx^{1-N}z
\nn \\
&&\left\{\left[\frac{P^2}{4}\left[y-y^2(x(1+z)-1)^2\right]
+q^2(1-y)y+P.py(1-y)(x(1+z)-1)-U(a)\right]^{-N}\right.
\nn \\
&&-\left.\left[\frac{P^2}{4}\left[y-y^2(x(z-1)+1)^2\right]
+q^2(1-y)y+P.py(1-y)(x(z-1)+1)-U(b)\right]^{-N}\right\} , \nn \\ && 
\eea
where we have defined
\be
U(a)=\left[\om x+a(1-x)\right]y+c(1-y) \, .
\ee
Making the substitution $x\rightarrow \tilde{z}$, such that
$\tilde{z}=x(1+z)-1$ in the second line of expression~(\ref{racek}), we can 
write for the first term in the large bracket~(\ref{racek})  (including all prefactors of the first line ~(\ref{racek})) 
\bea\label{konvert}
(-)\frac{\int_{\cal{S}}}{(4\pi)^2}
\int_0^1 dy\int_{-1}^z d\tilde{z}
\left(\frac{1+\tilde{z}}{1+z}\right)^{N-1}
\frac{z}{1+z}\frac{\rho_v(a)\rho_v(b)}{2(1-y)^N
\left[F(\tilde{\Omega},\tilde{z};p,P)\right]^N} \, ,
\eea
where $F$ is defined in Eq.~(\ref{efko}), and where
%page7pi
\be \label{konvert2}
\tilde{\Omega}=\frac{\left(\om\frac{1+\tilde{z}}{1+z}
+a\frac{z-\tilde{z}}{1+z}\right)y+c(1-y)
-\frac{P^2}{4}y^2(1-\tilde{z}^2)}
{y(1-y)} \, .
\ee
Introducing the identity 
\be
1=\int_0^{\infty}d \tilde{\om} \delta(\tilde{\om}-\tilde{\Omega})
\ee
into the expression~(\ref{konvert}), changing the integral ordering, and integrating over the
Feynman variable $y$, we arrive at the desired expression for ~(\ref{konvert})
\bea   \label{result1}
&&\int_0^{\infty} d \tilde{\om}\int_{-1}^1 d\tilde{z}
\frac{\chi_1(\tilde{\om},\tilde{z})}
{\left[F(\tilde{\omega},\tilde{z};p,P)\right]^N};
\nn \\
&&\chi_1(\tilde{\om},\tilde{z})=
-\int_{\cal{S}}\frac{T_+^{N-1}}{2(4\pi)^2}
\frac{z\rho_v(a)\rho_v(b)}{1+z}\theta(z-\tilde{z})
\sum_{j=\pm}\frac{y_{aj}\theta(D_a)\theta(y_{aj})\theta(1-y_{aj})}
{(1-y_{aj})^{N-1}\sqrt{D_a}} \, ,
\eea
where $y_{a\pm}$ are the roots of the quadratic equation
\be
y^2A+yB_a+c=0 ,
\ee
with the functions $A, B_a, D_a$ defined as
\bea \label{delfin}
&&A=\tilde{\om}-S; \hspace{1cm}
B_a=(\om-a) T_+ +a-c-\tilde{\om};
\hspace{1cm}D_a=B_a^2-4A c;
\nn \\
&&S=(1-\tilde{z}^2)\frac{P^2}{4};\hspace{1cm} 
T_{\pm}=\frac{1\pm\tilde{z}}{1\pm z}.
\eea
Similarly, we can repeat this whole procedure for the second term in
expression~(\ref{racek}), which gives
%
%page10pi
\bea \label{result2}
&&\chi_2(\tilde{\om},\tilde{z})=
\int_{\cal{S}}\frac{T_-^{N-1}}{2(4\pi)^2}
\frac{z}{1-z}\rho_v(a)\rho_v(b)\theta(\tilde{z}-z)
\sum_{j=\pm}\frac{y_{bj}\theta{(D_b)}\theta(y_{bj})\theta(1-y_{bj})}
{(1-y_{bj})^{N-1}\sqrt{D_b}},
\nn \\
&&y_{b\pm}=\frac{-B_b\pm\sqrt{D_b}}{2A};\hspace{1cm}B_b=(\om-b) T_-
+b-c-\tilde{\om}; \hspace{1cm}D_b=B_b^2-4A c. 
\eea
where we now use the label $\chi_2$ instead of $\chi_1$.

Next we transform  the last term in the second line of
expression~(\ref{prava2}) to the desired IR form~(\ref{repr}). For that
purpose, we basically follow the derivation already presented in
Ref.~\cite{SAUADA2003}, which is essentially the same. However, because of
the notational differences, we give all calculational details below.

Let us denote the relevant integral as
\bea \label{kralik}
&&I=i \dq \frac{1}
{D_1D_2D_3\left[F(\omega,z;q,P)\right]^{N-1}} .
\eea
Using the Feynman-parametrization technique, we first write 
\bea
  D_1 D_2&=&
 \frac{1}{2} \int\limits_{-1}^1 
 \frac{d \eta}{[M^2- f(q,P,\eta)- i \epsilon]^2} \, , \nonumber \\
  M^2\equiv \frac{a+b}{2}+ \frac{a-b}{2}\, \eta \,& ;&f(q,P,\eta)=q^2+\eta \, q.P +P^2/4 \, .
\eea
Then, the denominator of the IR of the bound-state vertex is added:
\bea
&&\frac{D_1 D_2}{[F(\om,z;q,P)]^{N-1}}=
\frac{\Gamma(N+1)}{2\Gamma(N-1)}\,  \int\limits_{-1}^1 d \eta
\int\limits_0^1 d t 
\frac{(1-t)t^{N-2}}{[R - f(q,P,\tilde{z})- i\epsilon]^{N+1}} \, , \nonumber \\
&&R= \om t + (1-t)M^2 \, , 
\eea
where $	\tilde{z}= t z+ (1-t)\eta$. Now, we match with the function $D_3$ 
and integrate over the four-momentum $ q$. Thus, we get 
\bea
&&I= i \,  \int \frac{d^4 q}{(2\pi)^4} 
 \frac{D_1 D_2 D_3}{[F(\om,z;q,P)]^{N-1}} \\
&&\hspace*{2.0truecm}
 =- i \, \frac{\Gamma(N+2)}{2\Gamma(N-1)}\,  \int\limits_{-1}^1 d \eta
\int\limits_0^1 d t\, (1-t)t^{N-2}\int\limits_0^1 d x\, x^{N}\, I_q \, ,
\nonumber\\
&&I_q= \int \frac{d^4 q}{(2\pi)^4} 
\left[ -q^2+ q \cdot Q - (1-x)p^2- \frac{x}{4}P^2
+(1-x) c+ x R- i\epsilon \right]^{-(N+2)} \nonumber \\
&& \hspace*{2.0truecm}
= \frac{i}{(4\pi)^2}\frac{\Gamma(N)}{\Gamma(N+2)}\frac{1}{x^{N}(1-x)^{N}}\,
\frac{1}{[ \Omega(t) - f(q,P,\tilde{z})- i\epsilon ]^{N}} \nonumber \\
&& \Omega(t)\equiv \frac{R}{1-x}+ \frac{c}{x}- \frac{x}{(1-x)}\, S \, ,
\eea
where $Q= (1-x)p- x \tilde{z} P/2$ and the function $S$ has been defined in
Eq.~(\ref{delfin}). Note that $\tilde{z}$ lies in the interval $[-1,+]>$ and
$0\leq S < (\sqrt{a}+ \sqrt{b})^2/4$. Interchanging the 
integrals over $\eta$ and $t$ such that
\bea && \int\limits_{-1}^{1}d\eta\int\limits_0^1 dt= 
\int\limits_{-1}^{1}d \tilde{z}\left[\int_0^{T_+}\frac {dt}{1-t}\, 
\Theta(z-\tilde{z}) +\int_0^{T_{-}}\frac {dt}{1-t}\, \Theta(\tilde{z}-z)\right] \, , 
 \nonumber\\ 
&&T_\pm= \frac{1 \pm \tilde{z}}{1 \pm z} \quad \quad \mbox{and} \quad \quad 
\tilde{z}= t z+ (1-t)\eta \, , 
\eea 
we finally obtain 
\be
I= \frac{N-1}{2(4\pi)^2}  \int\limits_{-1}^1 d \tilde{z}
 \int\limits_0^1 \frac{d x}{(1-x)^{N}} \sum_{s=\pm} \Theta(s(z-\tilde{z}))
 \int\limits_0^{T_s} \frac{d t\, t^{N-2}}{[F(\Omega(t),\tilde{z};p,P)]^{N}} \, .
\ee

Furthermore, we make the $t $dependence of $F(\Omega(t),\tilde{z};p,P)$
explicit as
\be
 F(\Omega(t),\tilde{z};p,P)= \frac{J(\om,z)}{1-x}\, t+
 F(\Omega(0),\tilde{z};p,P) \, ,
\nonumber 
\ee
where 
\bea
&&\Omega(t)=\frac{R(t)-S}{1-x}+\frac{c}{x}+S,
\nn \\
&&R(t)=J(\om,z)t+\frac{b+a}{2}-\frac{b-a}{2}\tilde{z},
\nn \\
&&J(\om,z)=\om-\frac{b+a}{2}-\frac{b-a}{2}z \, .
\eea
Integrating over the variable $t$ \/yields
\be
\int \frac{d t\, t^{N-2}}{ F(\Omega(t),\tilde{z};p,P)^{N}}= 
\frac{t^n}{(N-1)\, F(\Omega(0),\tilde{z};p,P)\,
[F(\Omega(t),\tilde{z};p,P)]^{N-1}} \, ,
\nonumber
\ee
and so
\bea \label{result3}
I&=&\frac{1}{2(4\pi)^2}
\int_{-1}^{1} d\tilde{z}\int_0^1 dx 
\sum_{s=\pm}\frac{\theta[s(z-\tilde{z})]T_s^{N-1}}
{(1-x)^NF(\Omega(0),\tilde{z};p,P)\left[F(A(T_s),\tilde{z};p,P)\right]^{N-1}} \, .
\nn \\
\eea
In order to separate the $F$'s in the denominator, we use the identity
\bea \label{op}
&&\frac{1}{F(\Omega(0),\tilde{z};p,P)\left[F(\Omega(t),\tilde{z};p,P)\right]^{N-1}}
\nn \\
&&=\frac{1-x}{J(\om,z)T_s}\left[\frac{1}{ F(\Omega(0),\tilde{z};p,P)}-
\frac{1}{F(A(T_s),\tilde{z};p,P)}\right]
\frac{1}{\left[F(A(T_s),\tilde{z};p,P)\right]^{N-2}} \, .
\eea
Note that, for a given $N$, one can repeat this algebra $N-1$ times 
till the  power of the last factor vanishes, which is precisely the reason to
use the trick~(\ref{op}). After this operation, the momentum dependence of the
denominator in each term becomes formally the same as in the desired IR. 
Although it is possible to derive the corresponding formula for arbitrary $N$,
this would lead to unmanageable expressions (probably not even in closed form),
so that we rather choose one concrete value of $N$.  Motivated by the success
of the scalar-model studies,  we take $N=2$ henceforth.
Explicitly, we obtain for $I$
\bea  \label{res4}
&&\frac{1}{2(4\pi)^2}
\int_{-1}^{1} d\tilde{z}
\int_0^1 dx \sum_{s=\pm}\frac{\theta[s(z-\tilde{z})]}
{J(\om,z)(1-x)}
\left[\frac{1}{ F(\Omega(0),\tilde{z};p,P)}-
\frac{1}{F(A(T_s),\tilde{z};p,P)}\right] .
\nn \\
\eea
Integrating by parts over the  variable $x$, we get for the latter expression
%page 14pi
%
\bea \label{resa5}
&&\frac{-1}{2(4\pi)^2}
\int_{-1}^{1} d\tilde{z}
\int_0^1 dx \sum_{s=\pm}\frac{\theta[s(z-\tilde{z})]\ln(1-x)}
{J(\om,z)}
\left[\frac{\frac{d\Omega(0)}{dx}}{ F^2(\Omega(0),\tilde{z};p,P)}-
\frac{\frac{dA(T_s)}{dx}}{F^2(A(T_s),\tilde{z};p,P)}\right] \, .
\nn \\
\eea
Implementing the identity
\be \label{arg}
1=\int_0^{\infty}\delta(\tilde{\om}-\Omega(t))
\ee
into the integrand of expression~(\ref{resa5}), changing the order of
integration, and carrying out the integration over the variable $x$, we
arrive at the desired result for the second term in the second line of 
expression~(\ref{prava2}):
\bea   \label{res5}
&&\int_0^{\infty} d \tilde{\om}\int_{-1}^1 d\tilde{z}
\frac{\chi_3(\tilde{\om},\tilde{z})}
{\left[F(\tilde{\omega},\tilde{z};p,P)\right]^N} \, ,
\nn \\
&& \chi_3(\tilde{\om},\tilde{z})=
\int_{\cal{S}}\frac{\rho_v(a)\rho_v(b)}{2(4\pi)^2}
\sum_{j=\pm}
\left\{\frac{\ln(1-x_j(0))}{J(\om,z)}
\theta[D(0)]\theta[x_{j}(0)]\theta[1-x_{j}(0)]
{\rm sgn}\left[\frac{dA(x_j(0))}{dx_j(0)}\right]\right.
\nn \\
&&-\left.\sum_{s=\pm}\theta[s(z-\tilde{z})]
\frac{\ln(1-x_j(T_s))}{J(\om,z)}
\theta[D(0)]\theta[x_j(T_s)]\theta[1-x_j(T_s)]
{\rm sgn}\left\{E[x_j(T_s)]\right\}\right\} ,
\eea
where we have included the previously omitted prefactor, and where
the $x_j$ are the roots of the delta-function argument in Eq.~(\ref{arg}),
viz.\
% page 16 pi
\bea
x_{\pm}(T)&=&\frac{-B(T)\pm\sqrt{D(T)}}{2A}
\nn \\
 D(T)=B(T)^2-4A c \, \, &&, \, \,
B(T)=R(T)-\tilde{\om}-c
\nn \\
\frac{d\Omega(t)}{dx}=\frac{E(x)}{1-x}
\, \, &&, \, \,
E(x)=\tilde{\om}-S-\frac{c}{x^2}\, . 
\eea
Introducing a more compact notation for the sum over an arbitrary function $U$ of
parameter $T$, namely
\be \label{konvence}
\sum_{\cal{T}}U(T)\equiv U(0)-\theta(z-\tilde{z})U(T_+)
-\theta(\tilde{z}-z)U(T_-) ,
\ee
we can rewrite $\chi_3$ in Eq.~(\ref{res5}) as
\be  \label{chi3}
\chi_3(\tilde{\om},\tilde{z})=
\int_{\cal{S}}\frac{\rho_v(a)\rho_v(b)}{2(4\pi)^2}
\sum_{\cal{T}}\sum_{j=\pm}
\frac{\ln(1-x_j(T))}{J(\om,z)}
\theta[D(T)]\theta[x_{j}(T)]\theta[1-x_{j}(T)]
{\rm sgn} E[x_j(T)].
\ee

The first term in the second line of expression~(\ref{prava2}) can be treated
in a very similar fashion as the previous case, though accounting for the
different power of $F$ in the denominator. Doing so explicitly, and multiplying
with the correct prefactor, the resulting expression reads
\bea \label{chi4}
&&\chi_4(\tilde{\om},\tilde{z})=
\frac{\int_{\cal{S}}}{2(4\pi)^2}
\left[\rho_v(a)\rho_v(b)\left(\om-\frac{P^2}{2}+\frac{a-b}{2}z\right)
-\rho_s(a)\rho_s(b)\right] \times 
\nn \\
&&\sum_{\cal{T}}\sum_{j=\pm}
\frac{\theta[D(T)]\theta[x_{j}(T)]\theta[1-x_{j}(T)]}{J^2(\om,z)}
\left\{\frac{TJ(\om,z)}{(1-x_j(T))|E[x_j(T)]|}
-\ln(1-x_j(T)){\rm sgn} E[x_j(T)]\right\}.
\nn \\
\eea
Assuming validity of the Uniqueness Theorem, we have converted the momentum BSE
into an equation for the weight function. It reads
\be  \label{eqvahy}
\rho^{[2]}(\tilde{\om},\tilde{z})=
\int_{0}^{\infty} d\om \int_{-1}^{1}dz
V^{[2]}(\tilde{\om}, \tilde{z};\om,z)\rho^{[2]}(\om,z) \, ,
\ee
where the kernel is simply given by the sum of the contributions derived above:
\be
V^{[2]}(\tilde{\om}, \tilde{z};\om,z)=
-3g^2 \int_0^{\infty}dc\int_0^{\infty}da\int_0^{\infty} db  
\rho_g(c)\sum_{i=1}^4\chi_i(\tilde{\om}, \tilde{z};\om,z).
\ee

\subsection{Heavy quark approximation --- unequal-mass case}

When the quark is sufficiently heavy (say $m_q(\mbox{2 GeV})>>
\Lambda_{\mbox{\scriptsize QCD}}$), then the approximation of the quark-mass
function by a constant turns out to be adequate. Neglecting self-energy
corrections is equivalent to the use of a free heavy-quark propagator,
which corresponds to the free-particle spectral functions
\bea
&&M_{1}\rho_v(a)=\rho_s(a)=M_{1}\delta(a-M_{1}^2) \, ,
\nn \\
&&M_{2}\rho_v(b)=\rho_s(b)=M_{2}\delta(b-M_{2}^2) \, ,
\eea   
where the variables $a$, $b$ distinguish the types of quarks the bound state
is composed of. For completeness, we write the kernel down explicitly for
this case:
\be
V^{[2]}(\tilde{\om}, \tilde{z};\om,z)=
\frac{-3g^2}{2(4\pi)^2} \int_0^{\infty}dc
\rho_g(c)\left(\chi_1+\chi_2+
\chi_3+\chi_4\right);
\ee

\bea
&&\chi_1=
-T_+\frac{z\theta(z-\tilde{z})}{1+z}
\sum_{\pm}\frac{y_{a\pm}\theta(D_a)\theta(y_{a\pm})\theta(1-y_{a\pm})}
{(1-y_{a\pm})\sqrt{D_a}},
\nn \\
&&\chi_2=
T_-\frac{z\theta(\tilde{z}-z)}{1-z}
\sum_{\pm}\frac{y_{b\pm}\theta{(D_b)}\theta(y_{b\pm})\theta(1-y_{b\pm})}
{(1-y_{b\pm})\sqrt{D_b}},
\nn \\  
&&\chi_3=
\sum_{\cal{T}}\sum_{j=\pm}
\frac{\ln(1-x_j(T))}{J(\om,z)}
\theta[D(T)]\theta[x_{j}(T)]\theta[1-x_{j}(T)]
{\rm sgn} E[x_j(T)],
\nn \\ 
&&\chi_4=
\left(\om-\frac{P^2}{2}+\frac{M_1^2-M_2^2}{2}z
-M_1M_2\right) \times 
\nn \\
&&\sum_{\cal{T}}\sum_{j=\pm}
\frac{\theta[D(T)]\theta[x_{j}(T)]\theta[1-x_{j}(T)]}{J^2(\om,z)}
\left\{\frac{TJ(\om,z)}{(1-x_j(T))|E[x_j(T)]|}
-\ln(1-x_j(T)){\rm sgn} E[x_{j}(T)]\right\} .
\nn \\
\eea
Here, the arguments $a,b$ of the functions $x,J,D$ must be replaced by the
quark masses $M_1, M_2$, respectively.

\subsection{Equal-mass case}

In the case of quarkonia, the kernel becomes more symmetric with respect to
the variable $z$, so that the formula for the kernel further simplifies.
The function $R$ depends on $z$ only through the variable $T$, such that
\be
J(\om,z)\rightarrow J=\om-M^2;
\hspace{2cm}
R(T)\rightarrow R(T)=J\,T+M^2 ,
\ee
where $M$ is the common mass $M=M_1=M_2$. The roots then become
\be
y_{a\pm}  \rightarrow x_{j=\pm}(T_+); \hspace{2cm}
y_{b\pm}  \rightarrow x_{j=\pm}(T_-)\, ,
\ee
and the kernel can be written in a more compact form:
\be
V^{[2]}(\tilde{\om}, \tilde{z};\om,z)=
\frac{-3g^2}{2(4\pi)^2} \int_0^{\infty}dc
\rho_g(c) K^{[2]}(\tilde{\om}, \tilde{z};\om,z,c) ;
\ee
\bea \label{equalmass}
&&K^{[2]}(\tilde{\om}, \tilde{z};\om,z,c)=
\nn \\
&&\sum_{s=\pm}\sum_{j=\pm}\theta[s(z-\tilde{z})]
\theta[D(T_s)]\theta[x_{j}(T_s)]\theta[1-x_{j}(T_s)]
\frac{-szT_s x_{j}(T_s)}{(1+sz) \sqrt{D(T_s)}(1-x_{j}(T_s))}
\nn \\
&&+\sum_{\cal{T}}\sum_{j=\pm}
\theta[D(T)]\theta[x_{j}(T)]\theta[1-x_{j}(T)] \times
\nn \\
&&\left\{\frac{(1-\frac{P^2}{2J})T}{|E[x_j(T)]|(1-x_j(T))}
+\ln(1-x_j(T)){\rm sgn} E[x_j(T)]
\frac{P^2}{2J^2}\right\} .
\eea

\subsection{One-gluon-exchange approximation}

In order to consider the one-gluon-exchange approximation, we must take the
massless limit $c\rightarrow 0$ and  restrict ourselves to the equal-mass case.
One can easily recognize that the root $x_+=0$ becomes trivial and the
associated contribution in expression~(\ref{equalmass}) vanishes.

For the purpose of brevity, we label
\be 
x(T)\equiv x_-(T)=\frac{\tilde{\om}-R(T)}{\tilde{\om}-S}.
\ee
Taking into account the relations
$$\sqrt{D}/x=A, \;\,\; E=A,$$
and doing a little algebra, we get for the  kernel
\bea \label{onefoton}
V^{[2]}_{\mbox{\scriptsize OGE}}(\tilde{\om}, \tilde{z};\om,z)&=&\frac{-3g^2}{(4\pi)^2} 
\left\{\frac{P^2}{4J^2}\theta(\tilde{\om}-m^2)\ln\left(\frac{m^2-S}{\tilde{\om}-S}\right)
\right.
\nn \\
&+&\left.\sum_{s=\pm}\theta[s(z-\tilde{z})]\theta(-B_s)\theta(A+B_s)
\left[\frac{T_s\left(\frac{1}{2}-\frac{P^2}{4J}\right)}{A+B_s}
+\ln\left(1+\frac{B_s}{A}\right)\frac{P^2}{4J^2}\right]\right\}. \nn \\ &&
\eea
%
%where, as it is dictated by our adopted convention (\ref{konvence}) 
%for $\cal{T}$ in the Rel.(\ref{equalmass}) the explicit relation for $x(0)$
%was used in the first line of  (\ref{onefoton}).
For completeness, we recapitulate here the complete list of functions:
\bea
&&A=\tilde{\om}-S,
\hspace{1cm}
J=\om-M^2,
\hspace{1cm} S=(1-\tilde{z}^2)\frac{P^2}{4},
\nn \\
&&B_s=(\om-M^2)T_s+M^2-\tilde{\om}
,\hspace{1cm} 
T_{\pm}=\frac{1\pm\tilde{z}}{1\pm z}.
\eea 

\subsection{Case $N=1$ }

Repeating the derivation, but now for the parameter value $N=1$, we should
obtain the homogeneous equation
\be  
\rho^{[1]}(\tilde{\om},\tilde{z})=
\int_{0}^{\infty} d\om \int_{-1}^{1}dz
V^{[1]}(\tilde{\om}, \tilde{z};\om,z)\rho^{[1]}(\om,z) ,
\ee
where the kernel is given by the expression
\bea
&&V^{[1]}(\tilde{\om}, \tilde{z};\om,z)=\int_S\frac{\sum_{i=1}^4\chi_i}{2(4\pi)^2};
\nn \\
&&\chi_1=
-\rho_v(a)\rho_v(b)\frac{z}{1+z}\theta(z-\tilde{z})
\sum_{j=\pm}\frac{y_{aj}\theta(D_a)\theta(y_{aj})\theta(1-y_{aj})}
{\sqrt{D_a}},
\nn \\
&&\chi_2=
\rho_v(a)\rho_v(b)\frac{z}{1-z}\theta(\tilde{z}-z)
\sum_{j=\pm}\frac{y_{bj}\theta{(D_b)}\theta(y_{bj})\theta(1-y_{bj})}
{\sqrt{D_b}},
\nn \\
&&\chi_3=
\rho_v(a)\rho_v(b)\sum_{j=\pm}
\frac{\theta[D(0)]\theta[x_{j}(0)]\theta[1-x_{j}(0)]}
{E[x_j(0)]},
\nn \\
&&\chi_4=
\sum_{\cal{T}}\sum_{j=\pm}
\frac{\theta[D(T)]\theta[x_{j}(T)]\theta[1-x_{j}(T)]}{|E[x_j(T)]|J(\om,z)}
\left[\rho_v(a)\rho_v(b)\left(\om-\frac{P^2}{2}+\frac{a-b}{2}z\right)
-\rho_s(a)\rho_s(b)\right]. \nn \\ &&
\eea
Here we use the same notations and conventions as in the previous section.
The corresponding derivation is very straightforward and exactly repeats
the steps made for the case $N=2$, so we only add a few comments.
The expressions for $\chi_{1,2}$ have in fact been derived in the previous
section, as they were for general $N$ see Rel.(\ref{result1}) and Rel. (\ref{result2}).
 The expression for $\chi_4$ we adopt
from Ref.~\cite{SAUADA2003}, while the remaining function $\chi_3$
follows from conversion of the term with $F^0=1$, i.e.,

\be
\int\frac{d^4q}{(2\pi)^4}\frac{1}{D_1D_2D_3} \, ,
\ee
which is present in the corresponding momentum-space BSE.
The proper derivation is in fact easier than in the $N=2$ case, and the
result represents the basic PTIR for a scalar triangle with one leg 
momentum constrained such that $P^2<(M_1+M_2)^2$.

\section{ Numerical procedure}

In this appendix we describe the numerical procedure actually used for
obtaining the bound-state spectra. Because of the structure of the integral
equations to be solved, the treatment is similar to the procedure used in
Refs.~\cite{KUSIWI,SAUADA2003}. However, we do introduce a rather tricky
modification here, which improves the numerical stability when the mass
parameter $m_g$  is
too small as compared to the constituent mass. Details of this technical
difference are given below. But first we describe the numerical treatment
for the case  $m_g\simeq M$ ($M$ is a heavy quark mass).

Equation~(\ref{equalmass}) is a homogeneous linear integral equation whose
solution needs to be properly normalized. For that purpose, let us adopt
the auxiliary normalization condition
\be \label{adopt}
1=\int_{-1}^{1} dz\int_{0}^{\infty} d\omega \frac{\rho^{[2]}(\om,z)}{J^2}.
\ee
Then we find that the equal-constituent-mass BSE can be transformed
into the inhomogeneous integral equation
\be  \label{actual}
\rho^{[2]}(\tilde{\om},\tilde{z})=K^{[2]}_I(\tilde{\om}, \tilde{z})
+\int_{0}^{\infty} d\om \int_{-1}^{1}dz
K^{[2]}_H(\tilde{\om}, \tilde{z};\om,z)\rho^{[2]}(\om,z)
\ee
with the kernels
\bea \label{cer1}
K^{[2]}_I(\tilde{\om}, \tilde{z})=
\frac{-3g^2}{(4\pi)^2}\frac{P^2}{4} 
\sum_{j=\pm}\theta[D(0)]\theta[x_{j}(0)]\theta[1-x_{j}(0)]
\ln(1-x_j(0)){\rm sgn} E[x_j(0)],
\eea
\bea \label{cer2}
K^{[2]}_H(\tilde{\om}, \tilde{z};\om,z,c)&=&
\frac{3g^2}{2(4\pi)^2}
\sum_{j=\pm}\theta[D(0)]\theta[x_{j}(0)]\theta[1-x_{j}(0)]
\ln(1-x_j(0))\frac{{\rm sgn} E[x_j(0)]-1}{J}
\nn \\
&+&\frac{3g^2}{2(4\pi)^2}\sum_{s=\pm}\sum_{j=\pm}\theta[s(z-\tilde{z})]
\theta[D(T_s)]\theta[x_{j}(T_s)]\theta[1-x_{j}(T_s)] \times
 \nn \\
&&\left\{\frac{szT_s x_{j}(T_s)}{(1+sz) \sqrt{D(T_s)}(1-x_{j}(T_s))}+
\right.
\nn \\
&&\left.\frac{\left(1-\frac{P^2}{2J}\right)T_s}{|E[x_j(T_s)]|(1-x_j(T_s))}
+\frac{\ln(1-x_j(T_s))}{J}{\rm sgn} E[x_j(T_s)]
\frac{P^2}{2J}\right\}.
\eea
The definitions of all these functions have been given in the previous appendix;
for the readers' convenience we recapitulate them here:
\bea
&&x_{\pm}(T)=\frac{-B(T)\pm\sqrt{D(T)}}{2A}
\, ; \, \, 
D(T)=B(T)^2-4A m_g \, ;
\nn \\
&&A=\tilde{\om}-S \, ; \,  \, B(T)=R(T)-\tilde{\om}-m_g \, ;
\nn \\  
&&E(x)=\tilde{\om}-\frac{m_g}{x^2}-S 
\, ; \, \,  R(T)=JT+M^2 \, ; \, \,   J=\om-M^2 \, ;
\nn \\
&&S=(1-\tilde{z}^2)\frac{P^2}{4} \, ; \, \,
T_{\pm}=\frac{1\pm\tilde{z}}{1\pm z} \, .
\eea
Equation~(\ref{actual}) has actually been used for the numerical solution of
the case $m_g/M \simeq 1$.
 
The kernel in Eq.~(\ref{actual}) is free from running singularities owing
to the presence of $\Theta$ functions. However, in the case of a
massless-gluon kernel, we would have a singularity just on top of the
boundary. There, $J\rightarrow 0$ as $\omega$ approaches the
quark mass. We find that
this instability is avoided if we generate the inhomogeneous term in the
following manner. First, we add exactly zero, in the form 
\be
(f(\tilde{\om},\tilde{z})-f(\tilde{\om},\tilde{z}))
\frac{\rho^{[2]}(\om,z)}{\om^2} ,
\ee
to the r.h.s.\ of the original homogeneous BSE for the weight function $\rho$.
Then, solving the equation  
\be  \label{actual2}
\rho^{[2]}(\tilde{\om},\tilde{z})=f(\tilde{\om},\tilde{z})+
\int_{0}^{\infty} d\om \int_{-1}^{1}dz
\left[V^{[2]}(\tilde{\om}, \tilde{z};\om,z) -
\frac{f(\tilde{\om},\tilde{z})}{\om^2}\right]\rho^{[2]}(\om,z) ,
\ee
with the normalization condition 
\be 
1=\int_{-1}^{1} dz\int_{0}^{\infty} d\omega \frac{\rho^{[2]}(\om,z)}{\om^2},
\ee
is equivalent to the solution of the original BSE. 
As a suitable function, we choose one with the property
\be
f(\tilde{\om},\tilde{z})=1 .
\ee
We observe that this method is applicable for any positive value of $m_g^2$,
but is only slowly convergent when it is used for the previously discussed
case $m_g\simeq M$. We also note that, up to a small numerical error,
Eq.~(\ref{actual}) yields the same spectra for those $\alpha$'s and $m_g$'s
which allow both Eqs.~(\ref{actual},\ref{actual2}) to be numerically stable. 

Equation~(\ref{actual2}) is solved by the method of iteration. When
straightforward iteration fails --- measure being  a difference between
weight functions obtained in different iteration steps, and/or a deviation of the
auxiliary normalization integral from a predefined value 
--- we change the coupling constant (in the treatment with fixed $P^2$,
otherwise the procedure is the converse) until a solution is found.
For the numerical solution we discretize the integration variables $\omega$ 
and $z$ using Gauss-Legendre quadrature, with a suitable mapping 
$(-1,+1) \rightarrow (\om_{\mbox{\scriptsize min}},\infty)$ for $\om$.
Equations~(\ref{actual2},\ref{actual}) are solved on a grid of $N=N_z*N_{\om}$
points spread over the rectangle
$(-1,+1)\times(\om_{\mbox{\scriptsize min}},\infty )$. The value
$\om_{\mbox{\scriptsize min}}$ is given by the support of the spectral
function. In all cases we take $N_{\om}=2N_z$.
Examples of numerical convergence for some bound-state cases are presented
in Table~3. As we can see, there is a rather weak dependence of the eigenvalue
$\alpha$ on the number of mesh points $N_{\om}$. The last column gives values
calculated from a weighted average (WA), with $N_{\om}$ the appropriate weight.

%Table 2
\begin{center}
\small{\begin{tabular}{|c|c|c|c|c|c|}
\hline
\hline
 $N_{\om}:$  &  32  & 40  & 64 & 80 & WA \\
\hline
\hline
$\eta=0.95;\, m_g/M=10^{-3} $ & 0.6611 & 0.6690 & 0.6697 & 0.6734 & 0.669\\
\hline
$\eta=0.95;\, m_g/M=0.5 $ & 1.037 & 1.0259 & 1.0210 & 1.0229 & 1.029\\
\hline
$\eta=0.99;\, m_g/M=0.5 $ & 0.818 & 0.8127 & 0.8158 & 0.8155 & 0.816\\
\hline
\hline
\end{tabular}}
\end{center}

\begin{center}
TABLE 3. The coupling $\alpha_s=g^2/(4\pi) $ for the ladder BSE with fixed
ratio $m_c/M$, as a function of the number of  mesh points. 
\end{center}

\mbox{ }

\end{document}